# DIMES: Let the Internet Measure Itself

Yuval Shavitt

Eran Shir

*Abstract*— Today's Internet maps, which are all collected from a small number of vantage points, are falling short of being accurate. We suggest here a paradigm shift for this task. DIMES is a distributed measurement infrastructure for the Internet that is based on the deployment of thousands of light weight measurement agents around the globe.

We describe the rationale behind DIMES deployment, discuss its design trade-offs and algorithmic challenges, and analyze the structure of the Internet as it seen with DIMES.

## I. INTRODUCTION

As the Internet evolved rapidly in the last decade, so has the interest in measuring and studying its structure. Numerous research projects [1], [2], [3], [4], [5], [6], [7], [8], [9] have ventured to capture the Internet's growing topology as well as other facets such as delay and bandwidth distributions, with varying levels of success. As the Internet continues to grow, especially far from its North American based core, measurement discrepancies are growing as well. A main handicap of current measurement projects is their rather limited number of measurement nodes (usually a few dozens up to a few hundreds) causing results to exhibit bias towards the core. In order to remedy this situation, a measurement infrastructure must grow several orders of magnitude in size and global dispersion.

We present DIMES, a highly distributed, global Internet measurement infrastructure, with the aim of measuring the structure and evolution of the Internet using a large set of interacting measurement agents. The key shift suggested in DIMES is the move from a small set of dedicated nodes, with measurements as their virtually sole objective, to a large community of host nodes, running light weight low signature measurement agents as a background process. Given the importance of location diversity in Internet measurements, this shift promises to enhance measurement results considerably.

Our goal is to map the Internet at several levels of granularity. At the coarse level, where each node is an AS, there are several mapping efforts, most notably are the active measurement Skitter project [5] and the passive collection of BGP data done by the RouteViews project [10], but also many of the studies mentioned above [1], [3], [6], [8], [9], [4], [7] examine the Internet (entirely or mostly) at this level. In the fine grain level, where each node represents a router, the mapping task is far more challenging, and the results achieved up to now [11], [2] are far from being satisfying. We believe that for many purposes neither of these granularities is appropriate. AS is too coarse a measure, where a node can represent a network that spans a continent or a small metropolitan ISP, while the router level is too fine to achieve a reasonable accuracy. Thus, our goal is to generate, on top of the other two maps, a mid-level granularity map where each node represents a group of routers working together, such as a small AS or a PoP of a large or medium size AS, like was suggested in the RocketFuel project [7].

## II. MOTIVATION FOR DIMES

Measuring the structure of the Internet is a daunting task. The Internet is a highly complex, evolving system. Routing between ASes in the Internet is governed by the Border Gateway Protocol (BGP) and its characteristics dominate the ability to reveal details about the AS interconnection. While it is outside the scope of this paper to explain BGP in details we will give one typical example that demonstrates how BGP disrupts our ability to reveal the Internet AS topology.

BGP is a path distance vector protocol, i.e., each AS announce to its neighbors not only the cost of its path to every destination but also the path itself. BGP is designed to enable Internet service providers (ISPs) to control the flow of data, thus an AS may choose not to announce some paths it knows due to *policy* which is determined by financial considerations. Our example is comprised of two small ISPs in the same geographic location in the Middle East. While these ISPs have BGP connections to some larger providers say one in North America and one in Europe (these large ISPs serve as their providers) they may wish to have a link connecting them directly to allow for low delay connection between their clients. However, none of the two local ISPs would like to serve as a transit AS for traffic destined to its peer since it has no financial incentive for this. For this reason BGP allows the ISPs not to broadcast to its providers its path to its peer AS. As a result, a researcher collecting BGP announcements from a point outside of the two local ISPs cannot learn about the existence of the local connection. An attempt to learn about this local peer-to-peer connection using traceroute from few measurements points will fail as well from the same reason. Only a presence in, at least, one of the two local ISPs will reveal the peer-to-peer link existence.

Previous studies [6], [4] show that by adding more vantage points, new links are revealed, and that the marginal utility of adding new vantage points decreases fairly fast. What escape these findings is the fact that while the marginal utility decreases, the mass of the tail is very significant, thus if one is using a few vantage points, say up to a few tens, there is a small advantage to add a few more, but there is a significant advantage to add additional thousands of points as they will add a significant percentage of new links. Using only a few dozens or even a few hundreds vantage points gives a strong bias in the topology to customer-provider links and misses many of the peer-to-peer links.

An additional problem with measuring the Internet in a non-local manner is that such measurements are non-efficient and non-friendly, redundantly traversing and measuring some of the edges thousands of times. Some smart algorithms [12], [13] were recently suggested to alleviate this problem.

These and other reasons make the case for a distributed, global, large scale measurement infrastructure. However, engineering a dedicated infrastructure with thousands of measurement computers spread around the globe is a feat only the largest of corporations can accomplish. Thus, in order to accomplish such a task, one must move to a distributed hosting paradigm, where lightweight measurement software is hosted by volunteers on computers all over the globe. Recently, the effectiveness of this approach has been demonstrated by several projects [14], [15], [16], [17] in various contexts, mostly related to computation intensive tasks. For Internet measurements, the contribution of a distributed approach is in the location heterogeneity. Using this approach, one can envision gaining presence in thousands of ASes. This paradigm shift has a tremendous effect on the underlying design principles and constrains of the infrastructure.

III. ARCHITECTURE GOALS AND GUIDELINES

DIMES's distributed architecture has three major goals: global and ubiquitous presence, the ability to host simultaneous experiments, and maximum experimental flexibility. Next, we will present these goals and describe the guidelines which stem from each of them.

DIMES shares many of its goals with previous measurement projects, most notably with NIMI [18], [19], which was an attempt to deploy a large number of versatile measurement software environments. The project suffered from the need to manage these entities which suppressed scalability. The DIMES distributed paradigm parts from previous attempts exactly at this point, making it possible to effortlessly scale up almost indefinitely.

DIMES is built to provide an accurate, timely and comprehensive map of the Internet, in terms of topology, latency, routing, and possibly bandwidth in the future. From this objective the central goal of DIMES is derived, i.e., to achieve global presence, in at least one of each two ASes which harbor an edge. Ideally, one would wish to have a DIMES agent in all ASes, and most of the routable IP prefixes. Practically, this goal puts our host set size marker on the order of several tens of thousands.

In order to establish a sustainable large community of users the DIMES architecture must follow several guidelines: provide security, constrain its resource consumption, provide incentives, and stay transparent. In addition the system must be scalable to enable the management and coordination of a growing community, and thus be easily paralleled and enhanced with new features. Next we will describe the rational behind these guidelines and their influence on the DIMES design.

*a) Security:* Being a platform with high degree of flexibility and remote programming abilities poses several serious security risks, such as the potential of hijacking the platform to perform DDoS attacks. Thus, it is of outmost importance to guarantee that the DIMES infrastructure is secured. To do this we do not keep the agent database on our web server, thus even a successful penetration into our web server will not provide the infiltrator with data about our agents. The agents do not expose the host machine to attacks since all the communication between the agent and the server are initiated by the agent.

*b) Constrain network resource usage:* As a guest in someone's machine, the DIMES agent must be polite in the way it uses network resources. First, the network resources usage should follow the well established strategy of distributed computing projects, that is giving the agent the lowest priorities and freeing resources whenever other processes need them. However, in the case of DIMES this strategy is not sufficient, as much of its potential measurement traffic patterns (such as repeatable traceroute and ping packets) is also associated in many networks' security infrastructures as rogue traffic. Thus, while the host computer may have abundant network resources available for the DIMES agent, it is important to establish strict constrains, so to not cause red flags to be risen by the administrators of the network to which the host belongs.

*c) Incentives:* As a system dependent on the good will of people, it is crucial for the success of DIMES to establish incentives which will generate enough interest to achieve sustainability. DIMES will incorporate two types of incentives: user experience incentives and knowledge incentives. User experience incentives consist of features such as dynamic visualization of Internet map segments and competition for data contribution. Knowledge incentives will consist of relevant information to the users, such as reports on ISP performance measures, reports on ease of access to user's web resources from other parts of the globe, and interfacing with web browsers and P2P applications to help them to optimize.

*d) Transparency:* To ease privacy concerns, it is important to be as transparent as possible. Thus, the DIMES platform is poised as an open-source platform.

An additional goal is high flexibility in the agent, allowing the easy deployment of new types of measurements, and the ability to create complex experiments. To fulfill this goal DIMES implements a plug-in mechanism for secure, user-driven auto-update of the agent. Such a mechanism allows the easy implementation of new measurement modules and bug fixes. In addition, we are developing PENny, a scripting language implemented through an agent based interpreter. PENny will enhance usual interpreted scripting languages with strong capabilities to blocking and non-blocking operation timing. Implementing a scripting language allows creation of complex scripts, with codependent operations, branching, and loops. Timed operations allow measuring consistently Internet time dependent features such as latency and bandwidth, as well as study routing instability.

Our third goal is to enable simultaneous experiment deployment, both in the infrastructure level and in the single agent level. We envision DIMES being used by numerous researchers, each studying a different facet of the Internet, possibly using different measurement modules. We must recognize that agents differ by their capabilities (e.g., some may have ICMP blocked), reliability, and mobility, and thus should be matched to the experiments where they will be most useful. Thus, the agent, the management system, and the experiment planning software are fitted with the following features:

- The management platform maintain a queueing system for experiments, which keeps track of agents' work scripts in flexible levels of aggregation, implementing a many-to-many relationship between agents and work scripts. There are queues for the entire system, per agents residing in the same AS, and for each agent. The scripts are provided to the agents bottom up, first scripts that are specific to the agent, then to its affiliated AS and finally to the entire system.
- There is a clear separation between experiment planning and management. Due to security and complexity issues, a distributed project such as DIMES needs to have a centralized management system. However, as DIMES is open to be used by the entire research community, its planning tasks were separated from its management tasks. That means that researchers are able to plan their own experiments using data received from DIMES (e.g., agents profiles, previous experiment results) oblivious of each other. The management platform is responsible for executing the experiments in parallel.
- The agent's scheduling mechanism receives operation requests from each of its running experiments, prioritize, while rigidly constraining the usage of networking and local resources.

## IV. THE DIMES ARCHITECTURE

Given the above requirements and guidelines we designed DIMES by balancing between flexibility for future growth and development speed.

*a) Programing Language:* DIMES is mostly written in Java. The two main reasons to choose Java are its natural sandboxing and security mechanisms, and the ease of portability for different operating systems. However, since Java does not support raw sockets, one still need to implement modules such as traceroute in native C++. Here, the modular architecture of DIMES helped us contain the usage of native code to specific sub-elements.

*b) Data aggregation:* Decisions on data aggregation formats have long term consequences. To defuse these effects, DIMES uses the MySQL relational database (RDB) to store the measurement results and their summaries. Using an RDB, rather than some flat file or XML formats (as is done by many other projects) provides us with a sought for flexibility in exporting and analyzing the data. It provides us with a way to bypass entirely the question of standardize data format, as once in an RDB, it is very easy to export the data in any format. The low price of storage makes it possible to deploy such an approach even with large amounts of data, of the order of tens of terabytes.

```
n = 0
nightResults = 0
onTime
startTime local 06/05/04 01:00
   while(currTime < local 01:30) {
     nightResults += Ping(198.81.129.100)
     n++ }
m = 0
dayResults = 0
onTime
startTime local 06/05/04 13:00
   while(currTime < local 13:30) {
     dayResults += Ping(198.81.129.100)
     n++ }
return dayResults/m - nightResults/n
```

Fig. 1. PENny script example

*c) Communication protocol:* DIMES is using http and https as the communication protocol for data and control, respectively. In today's Internet, there are many networks where all other TCP based traffic is being blocked, making other options impossible.

*d) Timed Scripting:* PENny, the DIMES scripting language can be considered as a somewhat diluted version of Java, implementing all of the basic branching, regular expression, and loop management features. On top of that PENny has substantial support for timing operations, using three time bases: local, GMT, and relative. These time bases allow for both local synchronization (e.g., implementing a measurement which runs everywhere at midnight) and gross global synchronization (e.g., starting a measurement at the same instant all over the world). In addition it provides a fairly accurate time dependence between measurements belonging to the same experiment (perform ping every 1 hour). Timing commands can be either blocking or non-blocking. PENny also has support for IP arithmetic with support for abilities such as prefix handling, IP incrementing, and IP mask-based randomizing. Thus, scripts can be written in a much more concise form, replacing IP lists with loop based scripting. Lastly, PENny is extensible, easily registering new commands when their respective measurement modules are installed. It can also handle rich environments, where modules are installed in subsets of agents.

It is important to state that PENny is designed, keeping in mind that much of the scripts ran on DIMES will be created automatically, rather than manually, as it is much more efficient when one wishes to tailor specific scripts to hundreds and thousands of agents. As such, a researcher designing a DIMES experiment is provided with previous experiments results, agents profiles and with an API for deploying PENny scripts. Figure 1 illustrates a simple PENny script which aims to identify delay differences on a specific path between night and day time.

*e) Authentication:* To mitigate security worries, DIMES is using an authentication mechanism. All communication other than agent to server results transferring is both secured (through https), digitally signed and verified using a public key mechanism. This is especially important for the auto-update mechanism which transfer runnable code, and as such poses the largest security threat.

*f) Open source:* DIMES is distributed under an open source license, and its source code is freely provided. This is first and foremost in order to reassure any privacy worries that may arise due to its constant usage of network resources, but is also a part of the DIMES strategy of opening up its capabilities to a larger community. There are DIMES modules under development by other groups, which are using our web based configuration control to make integration easier.

### A. Processes

*1) Communication with agent:* In designing the agent-center communication one needs to choose between pull, push, or combined model. For sending command scripts and software updates, the most natural way is for the center to push data to the agent every time the center has a new script for the agent or a new version release. For data collection, the most natural implementation is for the agent to initiate communication when data is ready. The main advantage in having the communication initiated by the center is that it is easy to control the communication load on the server. However, since agents are not always activated it poses a heavy burden of managing lists of tasks per agent. In addition, having the agent listen on some known port makes them vulnerable to malicious hijack attacks. Lastly, agents may be behind firewalls that block any communication initiated from the outside. We thus selected to have the agents initiate all communication with the server. The inability to ask an agent to perform a specific script is mitigated by our experiment planning concept that assign tasks to groups of agents (say grouped by AS or address prefix) rather than individual agents (which is also supported). This forces us to track agent reliability, mobility, and capability and use this information in our experiment planning tool. To protect the privacy of our users, we maintain only the following necessary information:

- **reliability**: We classify the agent reliability on daily and weekly time scales. An agent is daily (weekly) reliable if it performed measurements in most of the recent days (weeks).
- **mobility**: We distinguish between stationary agents, that always measure from the same IP address prefix; bi-homed which measure from two IP address prefixes, and mobile. For all agents, we keep the two top IP prefixes where they measure from, since even a mobile agent is performing a high percentage of its measurements from one or two locations.
- **capabilities**: We track agents capabilities, such as ability to perform ICMP tracerute, UDP pings, etc.

*2) Data aggregation:* As mentioned above, data aggregation is a bottleneck in our design. To mitigate this we use a lightweight storing process for new incoming data files, and use a low priority background process to parse the files and insert their content into the database. In addition, the raw data is stored as well. There is a trade-off between processing data by the agent, or sending all of the raw data for processing at the center. Processing by the agents reduces the agent-center communication load and the center computation load. However, the raw data can be useful for identifying problems or features that were not part of the initial experiment planning. These decisions should be made in the context of the measurement module being deployed and the experiment type. The code in Figure 1 is an example were all the processing is done at the agent, however, currently our experiments brings most of the data back to the center.

When an experiment is concluded, its planner may request to receive the information, both for analysis and for subsequent experiment planning. This allows for the creation of a feedback loop for experiment optimization. Our future plan includes dynamic experiment planning where automated tools analyze the arriving results and produce new scripts 'on-the-fly'.

## V. ANALYZING DIMES PERFORMANCE

### A. Building the case for DIMES

The underlying claim of the DIMES approach is that for accurately measuring the Internet's topology one must abundantly use distributed measurement nodes. To establish this claim, we need to compare DIMES results to results coming from traditional approaches, showing a significant, qualitative difference, and to show that the agents' contribution distribution has a heavy tail, meaning that new agents added to the DIMES platform contribute a considerable amount of new information.

The Route Views project [10] gather BGP updates from about 70 BGP speakers around the world. The BGP updates gathered in Route Views are freely available to download and are the largest open passive measurement database. As such, AS topologies inferred from Route Views data are the best yardstick against which measurement projects should compare themselves, at least at the AS realm. Given the dynamical nature of the Internet, one should be careful in comparing topologies, making sure that the topologies relate to the same time period and scope. Thus, in order to appropriately compare the DIMES topology to BGP inferred topology, it was necessary to take an integration of BGP updates during the measurement period. Thus we sampled BGP updates from Route Views, choosing one BGP update per day. Interestingly, integrating all these updates added less than 20% additional AS edges and less than 5% additional AS nodes in the resulting Route Views data inferred topology (BGP topology). In fact, the number of edges disappearing/appearing from the BGP topology per month in the last nine months stands roughly on 1500 edges, with deviations of less than 10%. This property of BGP topology will be used in the future to identify the closeness of DIMES topology to steady-state, given that we expect a similar (albeit somewhat larger) monthly variance.

Unless mentioned otherwise, below we refer only to the undirected version of the topologies.

## B. Data Collection Methodology

Up to June $1^{st}$ 2005 we collected about seventy six million measurements consisting of about sixty million traceroutes and sixteen million pings, from over 3000 agents, spread in more than 350 ASes. Out of these sixty million traceroutes, one can build various AS topologies, but no matter what measurement set we take, the first step which needs to be done is to infer IP-AS relationships. In order to translate IP level paths provided from the traceroutes to an AS level topology, one needs to associate IP addresses to ASes. Our current approach for the association process is to mimic a router's decision making process using a longest prefix matching algorithm, which looks for the longest prefix in our database that matches the IP in question. The prefix database, in turn, is built from prefix announcements in BGP data. The resolution process is augmented with a second tier consisting of whois data resolution, which is performed for IP addresses for which the main process has failed. Typically about 2-3% of the IPs fail the longest prefix matching and are resolved using whois. Currently, between 1-1.5% of the IPs fail AS resolution entirely. The translation process is somewhat challenging due to several issues surveyed in [20], [21].

The current number of discovered AS edges since project initiation stands on more than $57000$ connecting some $15000$ ASes. However, to be on the conservative side, we have decided to analyze a topology which is based on measurements performed only between March $1^{st}$ and June $1^{st}$ and consider an AS edge only if it was found by at least two separate measurements.

## C. Comparing DIMES vs. BGP Topologies

In the following, we will compare four different topologies that were created from the set of measurements defined above: DIMES topology, which is the AS level topology inferred from the DIMES measurements set; BGP Topology, which is the topology inferred from BGP updates gathered from Route Views during the similar period of March $1^{st}$ to June $1^{st}$; Complete Topology, which is the unification of the DIMES and BGP Topologies; and BGPinDIMES Topology, which is the BGP Topology subgraph which spans only AS nodes that belong to the DIMES topology. Table I shows the main properties of the four topologies (in this order): the number of nodes and edges; the average node degree; the power law exponent in the degree distribution; and the clustering coefficient.

As one can see, the BGP topology has about 25% more nodes than the DIMES topology. This difference is due to two main reasons. The first reason is that many ASes (for example military ASes and some corporations ASes) block active probes such as traceroute with various methods. To circumvent this issue to a certain degree the new version of the DIMES agent, which was just recently deployed, is augmenting the ICMP traceroute, which we used in the previous versions, with UDP based traceroute. We also plan the introduction of TCP SYN probes in the near future. The second reason for this gap is due to a lack of destination coverage, where we have not identified yet IP addresses which we can measure to in these ASes.

| Topology | N | E | $<k>$ | $\gamma$ | CC |
|---|---|---|---|---|---|
| DIMES | 14196 | 39084 | 5.51 | -2.1704 | 0.598 |
| BGP | 20397 | 47959 | 4.7 | -2.319 | 0.416 |
| Complete | 20585 | 60355 | 5.86 | -2.1854 | 0.596 |
| BGPinDIMES | 13959 | 36009 | 5.16 | -2.2481 | 0.419 |

TABLE I

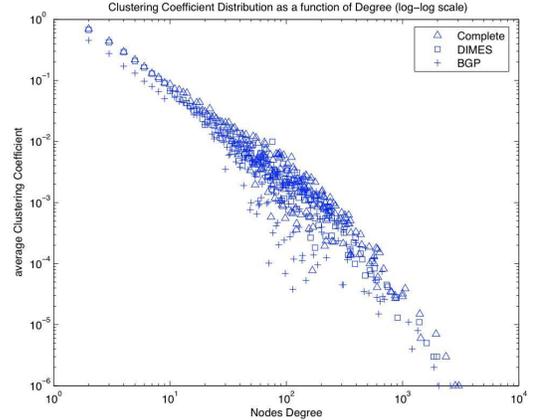

Fig. 2.

There are several conclusions that can be drawn from comparing these topologies. First, the degree distribution power exponent remains robust and hardly changes between the topologies, making it a poor characterizer of network differences. Thus, we should look for deeper topological characteristics to compare by [22].

When we compare the clustering coefficients of the DIMES topology and the BGP inferred topology, an immense difference of more than 40% appears. The difference cannot be attributed to the part of the network DIMES manage to measure since the clustering coefficients of the BGP and BGPinDIMES topologies is roughly the same, and it is well below the DIMES and COMPLETE figures, which are also roughly the same. In figure 2 we compare the clustering coefficient distribution of the topologies, showing the large difference in clustering coefficients of low degree nodes as well as the apparent under-sampling of middle degrees clustering in BGP. This property shows that many of the new links found by DIMES are periphery peer links. Finding these rich structures in the periphery was one of the main motivations for constructing DIMES.

Figure 3 compares the degree of nodes in the BGP topology vs. their degree in the Complete topology, namely with the DIMES contribution. We observe that about one third of the nodes has a higher degree than perceived by the BGP data.

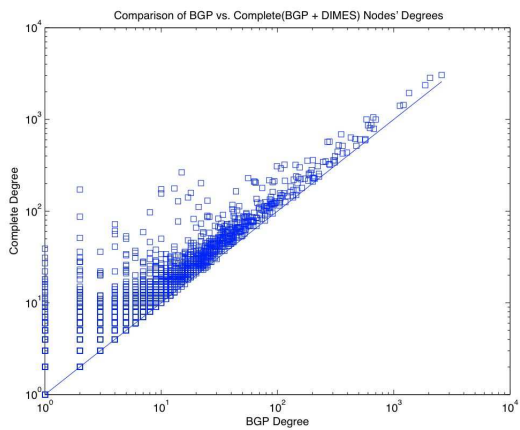

Fig. 3.

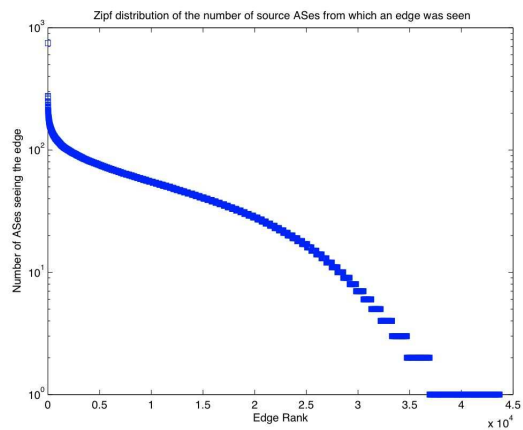

Fig. 4.

The highest degree in the remaining two thirds of ASes is 238, which means that all high degree nodes are augmented with edges, many of them to a considerable amount. AS 7018 (AT&T), for example, has more than 900 DIMES edges which do not appear in the integrated BGP topology, increasing its degree by more than 40%. DIMES data has contributed more than 500 new AS links to UUNet, Sprint, and Level3. This is in contradiction to the data presented by Chang *et al.* [23, Fig. 2] that claims that a few vantage points manage to reveal the complete connectivity of the highest degree ASes. Two possible reasons for the discrepancy are BGP rules that dictates some peering not to be announced, and peering without BGP, namely with static entries in the forwarding tables. However, focusing on the top 50 hubs of the network only tells portion of the story. The other 6400 ASes with BGP degree lower than 238 and DIMES edges not appearing in BGP constitute the bulk of the DIMES contribution. The lower degree nodes show sometime huge differences in degrees, which rise up to 30-fold and more.

An interesting question is how robust are the excess edges which do not appear in BGP, where by robust we refer to the number of ASes we see the edge from and the number of measurements that the edge was a part of. In figures 4 and 5 we present the edge count distribution as a function of number of ASes from which it was measured and as a function of number of measurements it belonged to, respectively. Remember, that the links with only a single measurements were removed from the DIMES topology; the remaining edges are mostly seen more than twice. We will discuss these graphs more in the next section.

### D. Agents contribution

The DIMES platform relies on volunteers enlisting into the system and installing the DIMES agent. As such, it is important to quantify their contribution, and specifically to quantify the contribution of new agents joining in the presence of many existing agents. Several authors [4], [6], [23] claimed that above a very low threshold (measured in few tens) additional agents' return will diminish and become unimportant. Looking into the contribution dynamics of the last nine months, one sees that the situation is far from it, as illustrated in figure 6. In this figure, the X axis represents days since project initiation, and the Y axis represents the ordered rank of the agents (i.e., agent who was 38 to join will have index 38). A point is plotted for each AS edge discovered according to the agent who discovered it and the date in which it was found. As can be seen, even agents that registered after tens of millions of measurements were performed still contribute substantially to the AS graph.

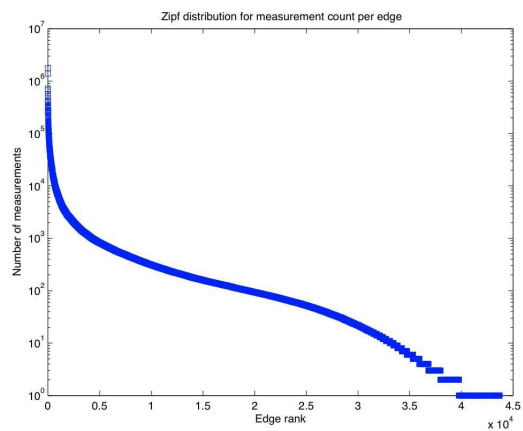

Fig. 5.

An interesting observation from figure 4 is that about 1/6th of the total DIMES edges have been seen only from a single AS, and additional 1/6th of the edges from 2-5 vantage points. Since we have measurements from only a few hundreds of ASes, we can assume that there are still many unknown edgeses we have not discovered, yet.[1]

---

[1] Indeed, additional data we have collected recently, which was not analyzed, yet, indicate that our recent measurement increased our Internet graph substantially.

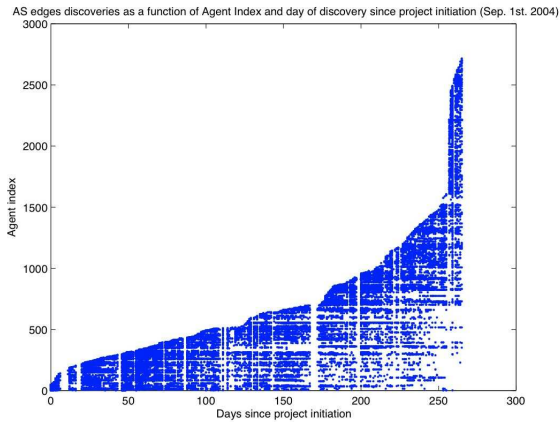

Fig. 6.

## VI. Data Analysis

In this section, we analyze and compare some of the properties of the AS, IP, and router level graphs. Let us first describe the way the various topologies were built. For the AS graph we have chosen the graph which was termed above "Complete" which is the union of the DIMES AS graph and the BGP inferred AS graph, both inferred from data collected between March and June. The IP graph is constructed directly from the traceroutes, where an edge is added for every pair of IPs which are adjacent in a traceroute path. It is important to note that there are many cases where routers do not send ICMP packets when they drop packets or do not answer ping echo requests. In such cases the router is identified not by its address (which is unknown in this case) but rather by a combination of his closest neighboring responding nodes. Specifically, it is defined by a triplet of two IP addresses and an index, where the index indicates the number of hops of the unknown router from the former closest neighbor which is responding (similar to [20]). However, in the current analysis, we investigate the subgraphs which consists of edges where both terminating nodes of each edge are associated with a known IP address.

The router graph is based on mapping one or more IP addresses (aliases) to a single router and then merging multiple edges between two routers. The current methodology of alias resolution which we deploy is based on performing a large scale UDP ping survey of all identified interfaces in our IP graph. When we send a UDP ping probe to IP address A from an agent a, the router will answer from an interface A' which is not necessarily equal to A, but in many cases is just associated with the router's interface which is closest to agent a or is the default responding interface of the router. This procedure is reproduced from many other agents distributed all over the world, a fact that increases substantially the possibility that no interface of the router will be left unresolved. We then group together IP addresses which are connected through a probe-response addresses couple and define them as a router. Up to June $1^{st}$ we have sent sixteen million ping probes from more than 3000 agents, out of which 3.7 millions were successful. Using these measurements we have managed to merge almost 80000 IPs that appear in the IP graph. While this is still far from being finished, it has already caused several significant differences between the IP and the router topologies.

In order to quantify how well our alias resolution process is, we compare routers' degrees to their respective alias rank, i.e., the number of IPs which are associated with the router. In figure 7 we present for each possible degree, the average number of aliases for all routers with that degree. While we are still missing many aliases, one can observe there is a correlation between the alias rank and the routers' degree. This comparison, in turn, can assist us in directing our efforts towards where we have the largest gaps in alias resolution.

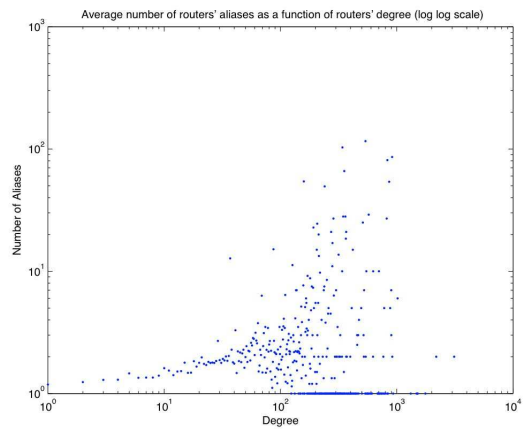

Fig. 7.

In table II we compare the properties of the IP, routers, and Complete AS graphs. Comparing the IP and routers graph we see that while the number of nodes has shrunken by less than 40000, the number of edges got smaller by almost 300000 edges, from which we can infer that many of the supposedly independent edges in the IP graph are in fact measurements of the same edge from different directions. The merging procedure had little impact on the degree distribution of the resulting router graph, but it had tremendous impact on other features of the graph, most notably its clustering coefficient and clustering distribution. Table II shows that the clustering coefficient of the router graph almost doubled compared to the IP graph. Investigating the entire clustering distribution in figure 8 we see that this ratio of two is kept almost throughout the entire distribution, signifying the importance of the alias resolution process. Given that the process is yet to be finished, we conjecture that quite possibly this ratio will increase as more interfaces are joined. The ongoing gap appears not only in the clustering distribution, but also when we investigate the average neighbor degree distribution. In figure 9 we observe that indeed the routers' distribution is above (though the gap is smaller) the corresponding IP distribution.

While indeed the clustering distribution for the routers is larger than the corresponding distribution for the IP graph by about two times through almost the entire scope of degrees, the two distributions are very similar in shape and structure. This cannot be said when we compare the AS distribution and the routers distribution. As we can see, again, from figure 8, the two distributions are very different qualitatively, especially in the low degrees where the mass of the graphs is situated. This difference can be stressed even more if we compare their corresponding average neighbor degree distribution (figure 9). For the most part, the routers' clustering distribution is confined to one order of magnitude, while the AS distribution is spread over two. In addition, while the AS neighbor degree distributions show clear dissasortativity, exhibiting a monotonic descend, the routers' distribution does not show the same characteristic and in fact in some range even shows an opposite behavior as observed and discussed previously [24], [25].

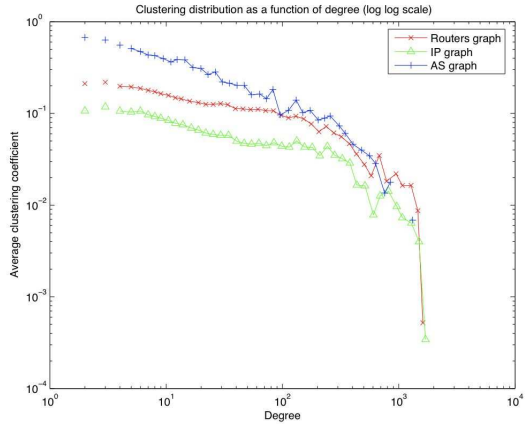

Fig. 8.

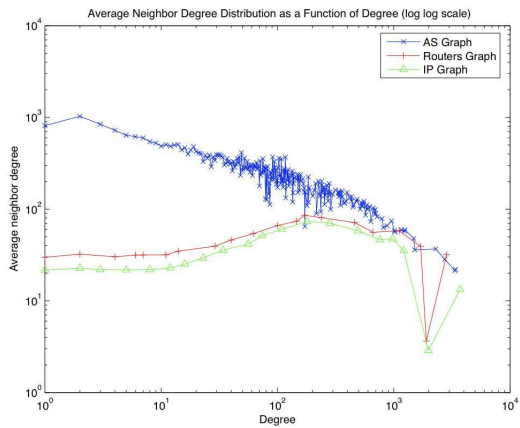

Fig. 9.

| Topology | N | E | $<k>$ | CC |
|---|---|---|---|---|
| IPs | 577127 | 2060079 | 7.1 | 0.097 |
| Routers | 539222 | 1774914 | 6.6 | 0.188 |
| AS-Complete | 20585 | 60355 | 5.86 | 0.596 |

TABLE II

Comparing the clustering coefficients of the router, and AS graph, we see that the AS graph has a clustering coefficient that is three times larger than the corresponding coefficient for the router graph. This large gap is maintained also for the entire distribution. This fact is somewhat in contrast to the fact that the AS graph is less, not more, dense than the routers graph. One potential explanation which we propose is hidden in the microscopic behavior of the nodes, mostly the low degree ones. We conjecture that two patterns govern small ISPs behavior. The first pattern consists of trying to reach as close to the core as possible. Thus, many ASes which have only two links to the world, will prefer to connect both these links to providers which reside in the core of the network. These providers, in turn, being in the core, have a high probability of being connected through a peering relation, which closes a triangle. The second pattern that is heavily deployed is the creation of local cliques or almost-cliques of peering relations between geographically neighboring ISPs. These constructs save the local providers resources, and thus are very frequent, increasing again the clustering rate for the low degree nodes. For the router graph, however, we claim that triangles (i.e., cycles of size three) have little routing significance, as the router network is a geographically localized network, and a triangle can be usually substituted better by a single router. Triangles do exist in the router topology, e.g., in PoPs where all the customer routers are connected to two backbone routers, but here one cannot find nodes that participate in hundreds of triangles like in the AS graph. Larger cycles will be relatively more prevalent in the router level graph since they save resources in connecting geographically distributed routers of the same AS.

In the AS graph there is a clearly distinguished core which functions as a hub that reaches all parts of the graph. In the IP and even in the router graphs, however, the core plays a relatively insignificant role, and in fact the highest degree nodes do not belong to the core of the graph. We elaborate this point by comparing the $k$-core [26], [27] decomposition of the graphs which is performed in the following way. In each step, starting with $k = 1$, we iteratively remove all the nodes whose degree in the residual graph is $k$ or below. The nodes which are removed at step $k$ belong to the $k$-shell, and the remaining nodes are the $(k+1)$-core. Plotting node degrees and shell sizes vs. shell index one identifies the source of difference between the AS and routers graphs.

In the AS graph (Figure 11), it is obvious that the graph has a very significant core, with nodes in the topmost shell having degrees which are more than 10 times higher than the degrees of the largest nodes in lower shells. Looking at the distribution of shell sizes it is clear that these high degree nodes must be heavily connected to the lower most shells (which we verified to be true), since the higher shells do not posses enough mass to account for the degree of the core nodes. The deep drop in the router and IP clustering coefficient graphs in figure 8 for the high degree nodes, is another indication to the lack of real core in these graphs. In [26] this issue is further investigated and it is shown that the AS topology is composed out of three distinct sections, one being the inner most core, the second being almost the entire first shell and the last being all of the middle shells, where the second and third segments are connected only through the first segment, namely, the inner core filling the topmost shell.

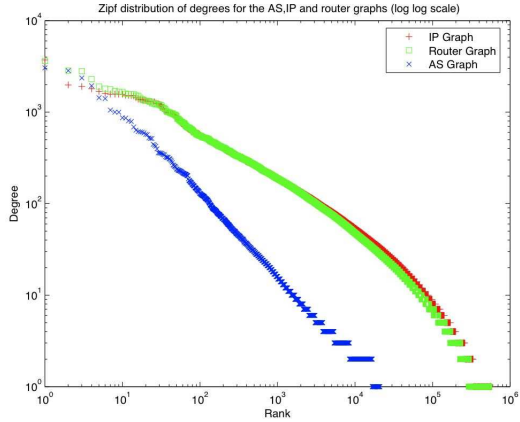

Fig. 10.

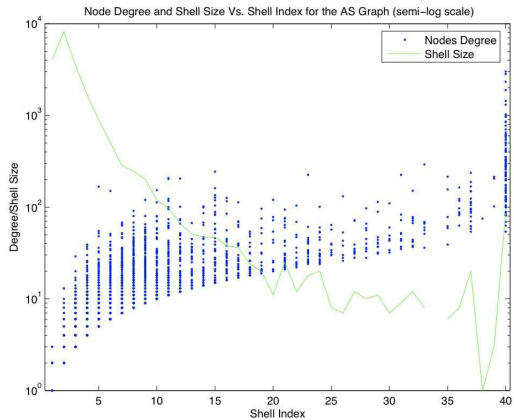

Fig. 11.

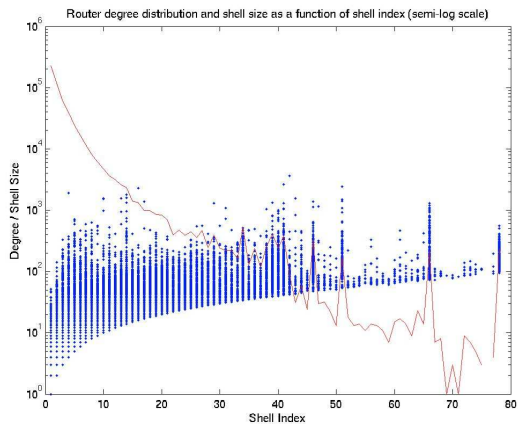

Fig. 12.

In the case of the router graph (figure 12), however, the situation is strikingly different. Two points can be easily observed. First, there are large gaps between shells 50 and 80, split by shell 66. Secondly, unlike the case of the AS graph, the highest degree nodes do not populate the topmost shell, but rather are spread over all shells. What can be inferred from this observed phenomena is that unlike the case of the AS graph, the top shells in the router graph do not play the role of central hubs, but rather are situated as bridges between a relatively small set of very large sub-networks.

## VII. CURRENT PROJECT STATUS

DIMES was launched on September 2004, and grew in its first nine months of operation to over 2300 users with over 3600 agents. The agents are spread in more than 70 countries mostly in N. America and Europe, but also in Asia, S. America, Africa, and the Middle East. Every day there are over 850 different agents performing measurements from over 350 different ASes. The current measurements rate performed by the DIMES platform exceeds 2.5 Million measurements per day, which entails a rate of more than a Billion measurements in the coming year.

The current version (Ver. 0.4) supports both ICMP and UDP traceroute, and ping. Future version will include TCP SYN probes. The agent supports Windows 2000 and XP, a Linux version will be released in the near future.

## VIII. FUTURE WORK

Looking further ahead, we intend to enhance DIMES in several directions. Obviously, we think of adding other types of measurements modules such as one for bandwidth estimation. We are planning to allow correlated measurements between two or more agents, and are looking at the security risks and technical difficulties (such as penetrating firewalls) that are associated with such tasks.

### ACKNOWLEDGEMENTS

We would like to thank Sorin Solomon and Scott Kirkpatrick for many insightful discussions; Anat Halpern and Ohad Serfati for their devotion in developing the DIMES software; and Amir Shay for his help in the initial stages of the project. The DIMES project is part of the EVERGROW integrated project which is supported by the EU 6th framework, IST Priority, Proactive Initiative Complex Systems Research. It is also supported by a grant from the Israel Science Foundation (ISF) center of excellence program (grant number 8008/03).

## REFERENCES

[1] M. Faloutsos, P. Faloutsos, and C. Faloutsos, "On power-law relationships of the internet topology," in *ACM SIGCOMM 1999*, Boston, MA, USA, Aug./Sept. 1999.

[2] R. Govindan and H. Tangmunarunki, "Heuristics for internet map discovery," in *IEEE Infocom 2000*, Tel-Aviv, Israel, Mar. 2000, pp. 1371–1380.

[3] L. Tauro, C. Palmer, G. Siganos, and M. Faloutsos, "A simple conceptual model for the internet topology," in *Global Internet*, Nov. 2001.

[4] P. Barford, A. Bestavros, J. Byers, and M. Crovella, "On the marginal utility of network topology measurements," in *ACM SIGCOMM IMW '01*, San Francisco, CA, USA, Nov. 2001.

[5] A. Broido and K. Claffy, "Internet topology: connectivity of IP graphs," in *SPIE International symposium on Convergence of IT and Communication '01*, Denver, CO, USA, Aug. 2001.

[6] Q. Chen, H. Chang, R. Govindan, S. Jamin, S. Shenker, and W. Willinger, "The origin of power-laws in internet topologies revisited," in *IEEE Infocom 2002*, New-York, NY, USA, Apr. 2002.

[7] N. Spring, R. Mahajan, and D. Wetherall, "Measuring ISP topologies with rocketfuel," in *ACM SIGCOMM '02*, Pittsburgh, PA, USA, Aug. 2002.

[8] A. Lakhina, J. W. Byers, M. Crovella, and P. Xie, "Sampling biases in ip topology measurements," in *IEEE INFOCOM '03*, San Francisco, CA, USA, Apr. 2003.

[9] S. Bar, M. Gonen, and A. Wool, "An incremental super-linear preferential internet topology model," in *PAM '04*, Antibes Juan-les-Pins, France, Apr. 2004.

[10] "University of Oregon Route Views Project," http://www.antc.uoregon.edu/route-views/.

[11] H. Burch and B. Cheswick, "Mapping the internet," *IEEE Computer*, vol. 32(4), pp. 97–98, 1999.

[12] B. Donnet, T. Friedman, and M. Crovella, "Improved algorithms for network topology discovery," in *PAM '05*, Boston, MA, USA, Mar./Apr. 2005.

[13] B. Donnet, P. Raoult, T. Friedman, and M. Crovella, "Efficient algorithms for large-scale topology discovery," in *ACM SIGMETRICS*, June 2005.

[14] "SETI@Home," http://setiathome.berkeley.edu/.

[15] "Distributed.net," http://www.distributed.net/.

[16] M. Dharsee and C. Hogue, "Mobidick: A tool for distributed computing on the internet," in *Heterogeneous Computing Workshop '00*, Cancun, Mexico, May 2000.

[17] J. Charles Robert Simpson and G. F. Riley, "Neti@home: A distributed approach to collecting end-to-end network performance measurements," in *PAM '04*, Antibes Juan-les-Pins, France, Apr. 2004.

[18] V. Paxson, J. Mahdavi, A. Adams, and M. Mathis, "An architecture for large-scale internet measurement," *IEEE Communications Magazine*, vol. 36, no. 8, pp. 48–54, Aug. 1998.

[19] V. Paxson, A. Adams, and M. Mathis, "Experiences with NIMI," in *PAM '00*, Hamilton, New Zealand, Apr. 2000.

[20] A. Broido and k. claffy, "Internet topology: connectivity of ip graphs," in *Proceedings of SPIE*, 2003.

[21] Z. Mao, D. Johnson, J. Rexford, and R. K. J Wang, "Scalable and accurate identification of as-level forwarding paths," in *INFOCOM*, 2004.

[22] L. Li, D. Alderson, W. Willinger, and J. C. Doyle, "A first-principles approach to understanding the internet's router-level topology," in *Proceedings of ACM Sigcomm 2004*, 2004.

[23] H. Chang, R. Govindan, S. Jamin, S. J. Shenker, and W. Willinger, "Towards capturing representative as-level internet topologies," *Computer Networks*, vol. 44, no. 6, pp. 737–755, Apr. 2004.

[24] M. E. J. Newman, "Mixing patterns in networks," *Phys. Rev. E*, vol. 67, no. 026126, 2003.

[25] R. Pastor-Satorras, A. Vázquez, and A. Vespignani, "Dynamical and correlation properties of the internet," *Phys. Rev. Lett.*, vol. 87, no. 258701, 2001.

[26] S. Carmi, S. Kirkpatrick, and E. Shir, "K-core analysis of the internet."

[27] J. I. Alvarez-Hamelin, L. Dall'Asta, A. Barrat, and A. Vespignani, "k-core decomposition: a tool for the visualization of large scale networks." [Online]. Available: http://arxiv.org/abs/cs/0504107